\documentclass[10pt,journal,twoside]{IEEEtran}
\usepackage{pdfsync}
\usepackage{epsfig}
\usepackage{epsfig}
\usepackage{amsmath}
\usepackage{amsthm}
\usepackage{amssymb}
\usepackage{comment}
\usepackage{cite}
\usepackage{graphicx}
\usepackage{lmodern}
\usepackage{colortbl}
\usepackage{multicol}
\usepackage{multirow}

\newcommand{\degree}{\ensuremath{^\circ}}

\begin{document}

\title{An Analytical Approach \\ for Memristive Nanoarchitectures}

\author{Omid~Kavehei,~\IEEEmembership{Student Member,~IEEE,}
        Said~Al-Sarawi,~\IEEEmembership{Member,~IEEE,}
        Kyoung-Rok~Cho,~\IEEEmembership{Member,~IEEE,}
		Kamran~Eshraghian, and~Derek~Abbott,~\IEEEmembership{Fellow,~IEEE}		
\thanks{Manuscript received September 22, 2011; accepted October 29,
2011. 
This work was supported by the World Class University project of MEST and KOSEF under grant no. R33-2008-000-1040-0 through Chungbuk National University and the Australian Research Council. 
}
\IEEEcompsocitemizethanks{\IEEEcompsocthanksitem O. Kavehei, S. Al-Sarawi, and D. Abbott are with the School of Electrical and Electronic Engineering, the University of Adelaide, SA 5005, Australia.\protect (email: omid@eleceng.adelaide.edu.au).
\IEEEcompsocthanksitem K.R. Cho and K. Eshraghian are with the College of Electronics and Information Engineering (WCU Program), Chungbuk National University, Cheongju, 361-763 South Korea.}
\thanks{Copyright~\copyright~2011 IEEE. Personal use of this material is permitted. However, permission to use this material for any other other purposes must be obtained from the IEEE by sending a request to pubs-permissions@ieee.org.}}

\markboth{MANUSCRIPT TO IEEE TRANSACTIONS ON NANOTECHNOLOGY. PREVIOUS ID: TNANO-00235-2011}{Kavehei \MakeLowercase{\textit{et al.}}: An Analytical Approach for Memristive Nanoarchitectures}

\IEEEcompsoctitleabstractindextext{%
\begin{abstract}
As conventional memory technologies are challenged by their technological physical limits, emerging technologies driven by novel materials are becoming an attractive option for future memory architectures. Among these technologies, Resistive Memories (ReRAM) created new possibilities because of their nano-features and unique $I$-$V$ characteristics. One particular problem that limits the maximum array size is interference from neighboring cells due to sneak-path currents. A possible device level solution to address this issue is to implement a memory array using complementary resistive switches (CRS). Although the storage mechanism for a CRS is fundamentally different from what has been reported for memristors (low and high resistances), a CRS is simply formed by two series bipolar memristors with opposing polarities. In this paper our intention is to introduce modeling principles that have been previously verified through measurements and extend the simulation principles based on memristors to CRS devices and hence provide an analytical approach to the design of a CRS array. The presented approach creates the necessary design methodology platform that will assist designers in implementation of CRS devices in future systems.

\end{abstract}

}

\maketitle

\IEEEdisplaynotcompsoctitleabstractindextext
\IEEEpeerreviewmaketitle

\section{Introduction}\label{sec:intro}
\IEEEPARstart{E}{merging} memory technologies based on new materials have been widely accepted as alternatives to the current CMOS technology. These technologies are mainly classified in three subclasses: Magnetoresistive Random Access Memory~(MRAM), Phase Change RAM~(PCRAM),  and Resistive Memory~(ReRAM)~\cite{ITRS:2009}. Memory applications motivate the need for an evaluation of these technologies in terms of READ and WRITE bandwidth, latency, and energy dissipation. The International Technology Roadmap for Semiconductors~\cite{ITRS:2009}, highlights that the performance characteristics of these emerging technologies are rather promising when compared with the curent large
memory arrays based on Static RAM (SRAM) constructs, particularly for large memory capacities. This suggests that emerging technologies, except PCRAM, will overtake advanced conventional Complementary Metal Oxide Semiconductor (CMOS) technology. We define a figure of merit as ${E_{\rm R}E_{\rm W}}{\tau_{\rm R}\tau_{\rm W}N_{\rm W,ref}}$ for comparing these technologies---with greater emphasis on the access time---the parameters represent READ and WRITE energy ($E_{\rm R}$, $E_{\rm W}$), READ and WRITE latencies ($\tau_{\rm R}$, $\tau_{\rm W}$), and the number of refresh cycles ($N_{\rm W,ref}$). This figure of merit indicates around $96\%$, $91\%$, and $79\%$ improvement for ReRAMs, MRAMs, and PCRAMs, respectively, over SRAMs for large memory capacities ($>1$ GB)~\cite{ITRS:2009}. The READ and WRITE access times of MRAMs show around $41\%$ more and $48\%$ less processing time than ReRAMs. ReRAMs introduce smaller cell size, $4F^2$/bit, where $F$ is the lithographic feature size, see Fig.~\ref{fig:memristor}(c), with comparable endurance in comparison with the other memory technologies. Although a number of strategies that utilize diode and/or transistor cross-point devices are proposed, their fabrication is relatively more complex than a ReRAM crossbar~\cite{fundamental:flocke:2007}.

The mathematical foundation of the memristor\footnote{The term {\it memristor} is a portmanteau of {\it memory} and {\it resistor}.}, as the fourth fundamental passive element, has been expounded  by Leon Chua~\cite{memristor:chua:1971} and later extended to a more broad class of memristors, known as memristive devices and systems~\cite{memristor:chua:1976}. This broad classification today includes all resistance switching memory devices such as ReRAMs~\cite{resistance:chua:2011}. Realization of a solid-state memristor in 2008~\cite{memristor:strukov:2008} has generated realization of both large memory arrays as well as new opportunities in the neuromorphic engineering domain~\cite{memory:pershin:2011, kavehei:2011:memristor-based, applications:ruhrmair:2010, read:csaba:2009, memristive:bushmaker:2010, integrated:xia:2011, field:ramanathan:2011}.

Although the memristor has introduced new possibilities for memory applications within the simple and relatively low cost crossbar array architectures, the inherent interfering current paths between neighboring cells of an addressed cell impose limitations on the scalability, a necessary condition for large memory arrays~\cite{crs:linn:2010}.

The imposed limitation was addressed by Linn~\emph{et al.}~\cite{crs:linn:2010} through adaptation of two series memristive elements connected with opposing polarities. This structure is referred to as Complementary Resistive Switch (CRS) as shown in Fig.~\ref{fig:crs_iv}. The unique aspect of this device is in using a series of high resistance states (HRS), $R_{\rm HRS}$, and low resistance states (LRS), $R_{\rm LRS}$, to introduce logic ``0" and logic ``1". As an example, a LRS/HRS combination represents ``1" and a HRS/LRS state represents ``0". Using this approach, the net resistance of the device is always around the HRS, $R_{\rm LOGIC}$, which helps in reducing sneak-path currents and at the same time main path currents. The advantage of using a CRS as a fundamental element originates from its excellent READ voltage margin, even with small HRS to LRS ratios. Moreover, it facilitates a comparable WRITE margin \cite{crs:yu:2010}. There is also a lack of SPICE model verification for CRS devices and a statistical analysis considering the mentioned operational uncertainties. Here, we address these issues using a Verilog-A implementation for memristor dynamics within a memristor macro-model for SPICE simulation. 

Contributions in this paper can be categorized in five parts:
\begin{itemize}
	\item CRS device modeling and verification using available functionality information from~\cite{crs:linn:2010}.
	\item Comprehensive mathematical framework for a memristive array for assisting designers to identify the impacts of stored memory pattern, parasitic resistances, and sneak-path currents on the array performance. In addition, we develop a system of linear equations to extract the voltage pattern across an array regardless of the READ or WRITE scheme. This mathematical framework can be used for different emerging memory devices.
	\item Characterization of a comprehensive framework for a CRS array and identifying an optimal value for load resistors in a CRS array.
	\item Highlighting the importance of the WRITE scheme and array size in the total power dissipation through an analysis of the number of half-selected cells in array for memristive-based and CRS-based arrays.	
	\item It is also shown that the existence of a long tail distribution in LRS resistance, after several thousand operation cycles, process variation, temperature effects, and uncertainties related to the nanowire parasitic resistances have significant impact on a memristor cross-point array than a CRS array. 
\end{itemize}

Along with addressing the parasitic current path issue, the importance of parasitic resistors also increases as $F$ reduces. In a practical memory design, the line resistance of a nanowire can be calculated with $R_{\rm line}=\rho_{\rm metal}({0.2n}/{F})$, where $n$ is the number of cells in the line and $\rho_{\rm metal}$ is the resistivity of metal, which is a function of $F$~\cite{memory:zhirnov:2010}. Therefore, a mathematical model is developed to consider the nanowire parasitic resistances in a matrix based analysis of cross-point arrays. 

In this paper, the preliminaries of memristor technology and modeling approach based on a fabricated Metal-Insulator-Metal (MIM) structure is presented in Section~\ref{sec:memristors}. Then a practical model of the CRS device is developed and created in SPICE as a macro-model. This is described in Section~\ref{sec:crs}. Section~\ref{sec:cross} shows the analytical cross-point model and mathematical model as well as simulation results and discussions. 

\section{Memristors}\label{sec:memristors}

Memristive device modeling is a necessary step for CRS device modeling. Therefore, we first present the memristive element characteristics, which can be defined using two equations,
\begin{eqnarray}
\left\{ \begin{array}{ll} \label{equ:memsystemequ}
I=g(w,V)\cdot V\\
\frac{{\rm d}w}{{\rm d}t}=f(w,V)~,\end{array}\right.\\ \nonumber
\end{eqnarray} 
where $w$ is a physical variable indicating the internal memristor state that in theory is such that $0<w<L$, where $L$ is the thickness of transient material oxide (TMO) thin-film. The parameters $I$ and $V$ represent current and applied voltage, respectively. The second expression of Eq.~(\ref{equ:memsystemequ}) defines velocity of this movement. Considering an ionic conduction mechanism, this part can be redefined as an ionic drift velocity. The function $f(\cdot)$ captures the highly nonlinear characteristics of the memristor as function of the applied voltage~\cite{coupled:strukov:2009,exponential:strukov:2009,switching:pickett:2009}.  
The $I$-$V$ curve relationship, as in Eq.~(\ref{equ:memsystemequ}), has already proposed by several groups~\cite{memristor:shin:2011, memristor:batas:2011, fabrication:kavehei:2011, feedback:yi:2011}, however, an accurate reproduction of the characteristics in simulation is an area of intense research.  
An appropriate $f(\cdot)$ function seems to be either a double exponential or related forms~\cite{switching:pickett:2009,fabrication:kavehei:2011}, or a $\sinh(\cdot)$ function~\cite{exponential:strukov:2009}, which defines intrinsic threshold voltages. Here we apply a commonly accepted $f(\cdot)$ function to our experimental data. The memristor state variable is identified through time integral of the $f(\cdot)$ function and then it is applied to the $g(\cdot)$ function. The outcome shows a good agreement between the measured data and the modeled $I$-$V$ hysteresis.

To address this modeling problem we use the Mott \emph{et al.}~\cite[Chap.~2]{electronic:mott:1964} model of ionic conduction in terms of the theory of lattice defects that has been already used in several studies in this area~\cite{crs:yu:2010}. In this case, 
\begin{eqnarray}
\frac{dw}{dt}=\upsilon_0e^{-\frac{U}{kT}}\sinh(\frac{\rho V}{kT})~,\label{equ:velocity}
\end{eqnarray}
where $\upsilon_0$ is initial velocity, $U$ is the potential barrier height, $k$ represents the Boltzmann constant, $T$ temperature, and $V$ applied voltage in eV. The parameter $\rho=a/2L$, which is a dimensionless parameter related to the distance between adjunct lattice positions, $a$, and TMO thickness $L$. For approximating the effective electric field in the equation, $\rho$ should be a function of $w$ in the way that $\rho=a/2(L-w)$. Quantities are summarized in Table~\ref{tab:params}.

The relationship between the drift velocity and device thickness, $\upsilon\propto \sinh(1/L)$, clearly shows the reason that memristive behavior appears strictly at nano scale dimentions. 
A detailed comparison between the model and experimental data is given by Heuer~\emph{et al.}~\cite{nonlinear:heuer:2005}.

The virgin device needs an electroforming step that acts like a soft breakdown condition and creates a conductive channel through the TMO material (TiO$_2$ in our case). Then the applied voltage polarity identify the channel orientation~\cite{switching:strachan:2011, electrochemical:valov:2011, nanofilamentary:kim:2011}. In our case, the forming step is carried out by applying an electric field around $6.2$~MV/cm across TiO$_2$. This forming step creates a difference in atomic percentage ratio of oxygen in TiO$_2$ close to one of the electrodes. The conducting mechanism is then carried out through a channel known as a conducting filament (CF). The conducting filament is highly localized (e.g. for a cylindrical CF, $A_{\rm CF}\approx 10$~nm$^2$) compared to the metalic contact area, $A$, and the filament (ON) resistance $R_{\rm ON}$ is proportional to $A^{-1}_{\rm CF}$~\cite{physical:ielmini:2011}. This shows that controlling the filament area is very important for controlling the programming current in an array of ReRAM elements. Controlling $A_{\rm CF}$ is possible through a set of operations. It is also reported that the SET threshold smoothly increases as $R_{ON}$ increases, $V_{\rm SET}\propto R^{0.25}_{ON}$, whereas a significant decrease in $I_{\rm SET}\propto R^{-0.75}_{ON}$ was reported~\cite{physical:ielmini:2011}.   

\begin{figure}[htb!]
\centering
  \includegraphics[width=0.45\textwidth]{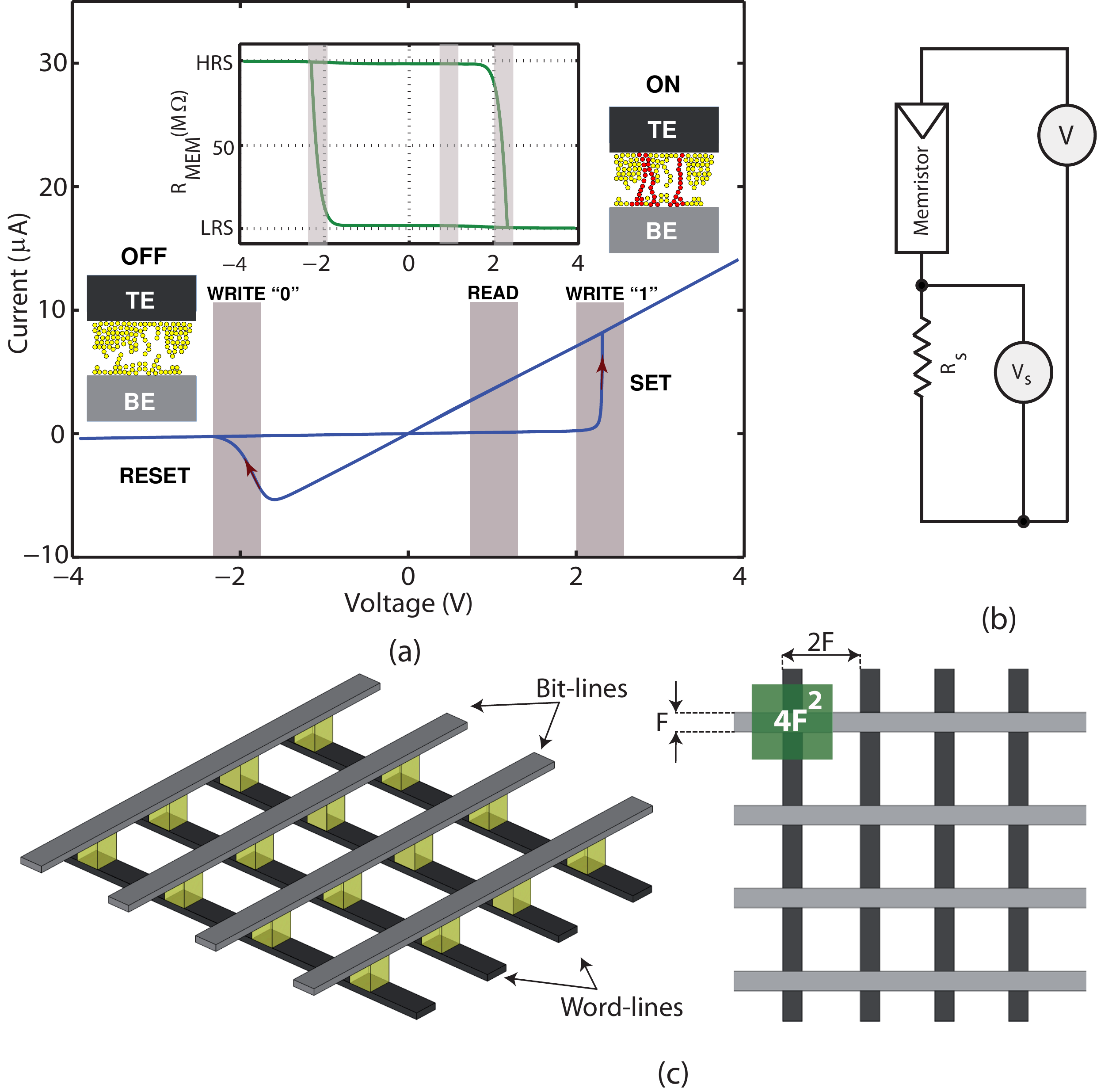}
  \caption{Memristor cross-point implementation and array presentation. (a) Memristor model result verified by experimental data of a fabricated Ag/TiO$_2$/TiO$_{2-x}$/ITO~\cite{fabrication:kavehei:2011}. Inset shows memristor device resistance {\it vs} applied voltage. TE and BE are stand for top-electrode and bottom electrode, respectively. The reason for choosing $-4$~V to $4$~V is related to CRS device functionality that is explained in Section~\ref{sec:crs}. The curve shows asymmetric characteristics. The measurement data was collected using a Keithley 4200 semiconductor characterization system. Red paths (inset (a)) show filament paths. The probability that a conductive path is broken can be calculated through a set of (independent) Boltzmann probabilities~\cite{memory:zhirnov:2010}. To avoid complexity, data for linear parts are not shown here. Due to the internal dynamics of the memristor, we applied similar voltage (triangular) signal to the model and device. The digram (b) conceptually shows the measurement setup. The arrays in (c) illustrate 3D and 2D representations of a ReRAM (memristive) array.}\label{fig:memristor}
\end{figure} 

A Verilog-A implementation of this model is used as a macro-model in Cadence. The macro-model implements the $g(\cdot)$ function as $I=(w/L)I_{\rm ON}+(1-w/L)I_{\rm OFF}$. Our experiments show that a linear $I_{\rm ON}$-$V$ is in agreement with measurement results for an area of $100\times 100~\mu$m$^2$~\cite{fabrication:kavehei:2011}. Further reduction of the device feature size, $F$, is possible. However, issue related to crosstalk and tunneling between neighboring cells should be taken into account \cite{memory:zhirnov:2010,resistive:akinaga:2010,burr:2010:phase}. The OFF current can be defined as $I_{\rm OFF}=\chi_1e^{(\gamma_1 V)}-\chi_2e^{(-\gamma_2 V)}$, where $\chi_1$, $\gamma_1$, $\chi_2$, and $\gamma_2$ are fitting parameters. Different $I$-$V$ equations for ON and OFF states have been also reported in~\cite{memristor:joshua:2008} and~\cite{nonvolatile:inoue:2005}. For instance, Inoue~\emph{et al.}~\cite{nonvolatile:inoue:2005} reported $I_{\rm OFF}\propto \sinh(\cdot)$ and $I_{\rm ON}\propto \sinh^{-1}(\cdot)$. Section~\ref{sec:sub:dis} includes a discussion on using the nonlinearity of the memristor characteristics for increasing $R_{\rm ON}$ at a given READ operation voltage.

\begin{table}[htb!]
  \centering
  \caption{Physical Parameters}
    \begin{tabular}{|c|r|l|c|} \hline \hline
    Parameter & Value & Units & Reference \\ \hline
    $a$     & $1.5$ & ${\rm \AA}$    & \cite{exponential:strukov:2009} \\ \hline
    $f_{\rm e}$     & $10^{13}$ & attempts/s & \cite{exponential:strukov:2009} \\ \hline
    $E_{\rm ai}$     & $1.1$ & eV    & \cite{observation:miao:2011} \\ \hline		
    $\upsilon_0$ & 1500  & m/s   & calculated \\ \hline
    $L$     & $22$ & nm    & fabricated \\ \hline
    $\rho$  & $0.0034$ & no units & calculated \\ \hline
    $k_{\rm th}$  & $1.5$ & W/(Km) & \cite{impact:xia:2011}  \\ \hline
    $A$ & $100\times 100$     & $\mu$m$^2$ & fabricated \\ \hline			
    $A_{\rm CF}$ & $10$     & nm$^2$ & \cite{physical:ielmini:2011} \\ \hline
    $R_{\rm th}$ & $4.5\times 10^{6}$   & K/W & calculated \\ \hline \hline
    \end{tabular}  \label{tab:params}
\end{table}

Fig.~\ref{fig:memristor} illustrates the modeling results for a memristor based on experimental data from a Silver/Titanium dioxide/Indium Thin Oxide (Ag/TiO$_{2}$/TiO$_{2-x}$/ITO) measurement implementation~\cite{fabrication:kavehei:2011}. This is a novel combination for an ReRAM implementation in using TiO$_2$ as the TMO material according to Table 1 in~\cite{resistive:akinaga:2010}. This implementation yields a bipolar cell with nearly $200$ successful cycles. Besides asymmetry, recent studies shows that the inherent Joule heating effect is responsible for (RESET) switching mechanism in a way that sufficient heat induces a crystallization of the oxide surrounding the channel~\cite{switching:strachan:2011, conductive:russo:2007}. This crystallization time frame is exponentially related to temperature~\cite{switching:strachan:2011}. The exact equation can be extracted from~\cite{physical:ielmini:2011}. Therefore, as Joule heating increases in the hysteresis, the CF diameter (hot spot) shrinks and this effect would lead to a reset. We include this effect in our macro-modeling approach using a relationship introduced in~\cite{filament:russo:2009},   
\begin{eqnarray}
T-T_0=PR_{\rm th}~,\label{equ:joul_heating}
\end{eqnarray}
where $T_0=300$~K, $R_{\rm th}=L/(8k_{\rm th}A_{\rm CF})$ is the thermal resistance, $P=IV$ is Joule dissipation at reset, and $k_{\rm th}$ is TiO$_2$ thin film thermal conductivity. According to experimental data, as $A_{\rm CF}$ decreases, RESET current decreases, so the RESET threshold voltage would increase~\cite{conductive:russo:2007}.

In order to increase simulation convergence Eq.~(\ref{equ:velocity}) can be rewritten as, 
\begin{eqnarray}
\frac{dw}{dt}=\upsilon_1V+\upsilon_3V^3+\upsilon_5V^5+\ldots~,\label{equ:newvelocity}
\end{eqnarray} 
where $\upsilon_1$, $\upsilon_3$, and $\upsilon_5$ are low-field and higher order coefficients. This approach also combines the effects of Joule heating and $L-w$ (on the effective electric field) in $\upsilon_i$ coefficients. Note that ${dw}/{dt}=\upsilon_1V$ usually defines a pure memristive behavior, as described in~\cite{memristor:chua:1971}.  

These properties then raise the following questions, (i) how to address the asymmetric characteristic in WRITE and READ operations, (ii) what is the impact of using more realistic model for cross-point array evaluation, and (iii) what is the effect of different device level $I$-$V$ characteristics on the array performance? Here we are aiming to answer the first two questions using the explained model and the answer to the third question is currently under review by the our research group and will be the topic of another paper. The second question can be answered using a worst-case consideration for $R_{\rm ON}$. In this case, this paper compares a memristor array with a CRS-based array.

Using the developed model of the memristor that accurately models the nonlinear behavior of the device, we can model the CRS as explained in the following section. 

\section{Complementary Resistive Switch}\label{sec:crs}

\begin{figure}[htb!]
\centering
  \includegraphics[width=0.4\textwidth]{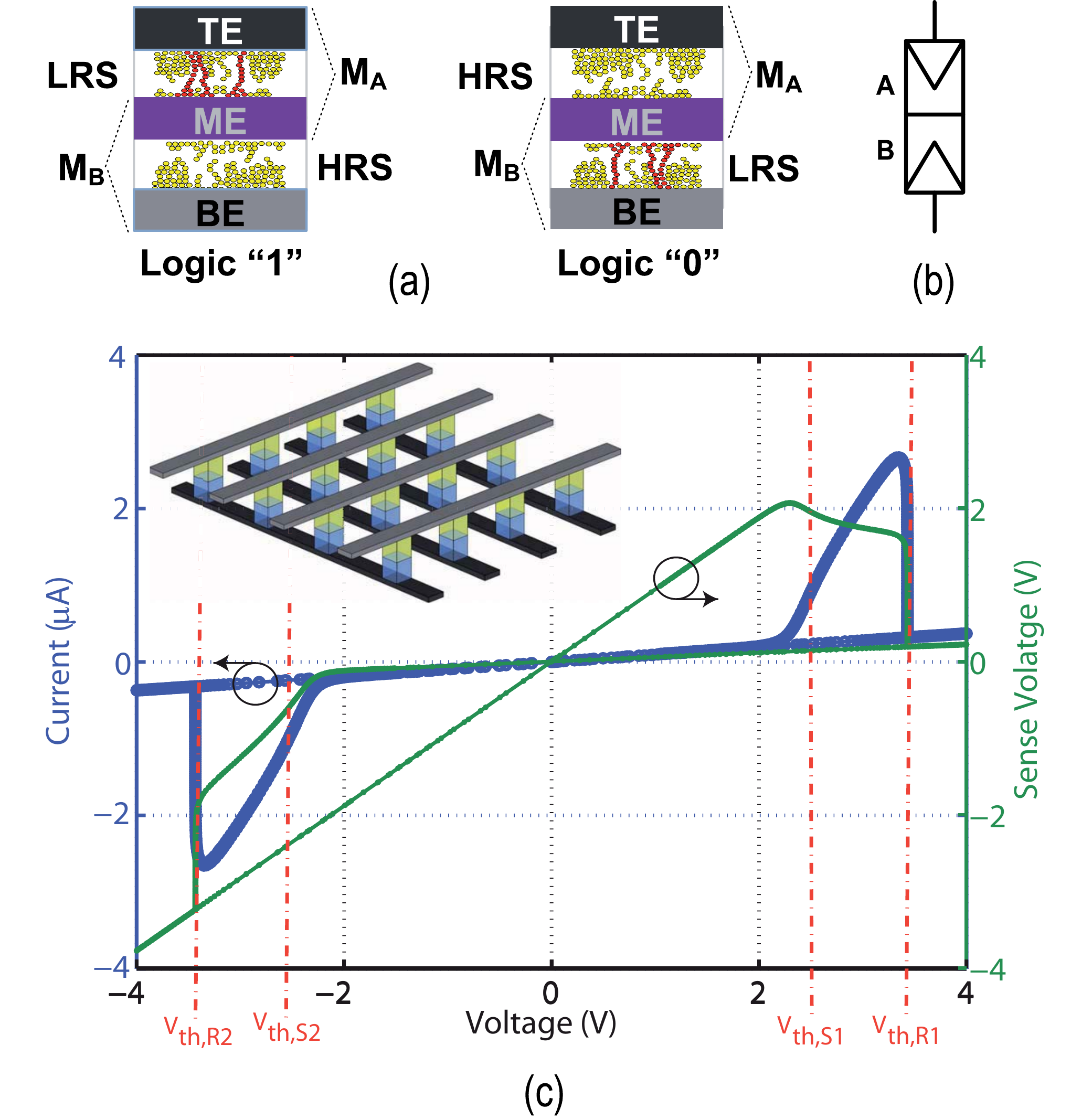}
  \caption{CRS functionality using the described memristor model. (a) Fundamental behavior of switching between logic ``0" and logic ``1". Here, TE, BE, and ME are the top, bottom, and middle electrodes, respectively. The reading procedure can be carried out by sensing ME. Note that M$_{\rm A}$ and M$_{\rm B}$ are memristors A and B. (b) A symbol for the CRS device. (c) $I$-$V$ sense voltage curve of the simulated device that shows the characteristics described in~\cite{crs:linn:2010}. The inset shows a 3D schematic of a CRS array. Note that the middle electrode (ME) should not be connected to any point, otherwise it creates new sneak current paths.}\label{fig:crs_iv}
\vspace{-22pt}
\end{figure}

\begin{figure}[htb!]
\centering
  \includegraphics[width=0.45\textwidth]{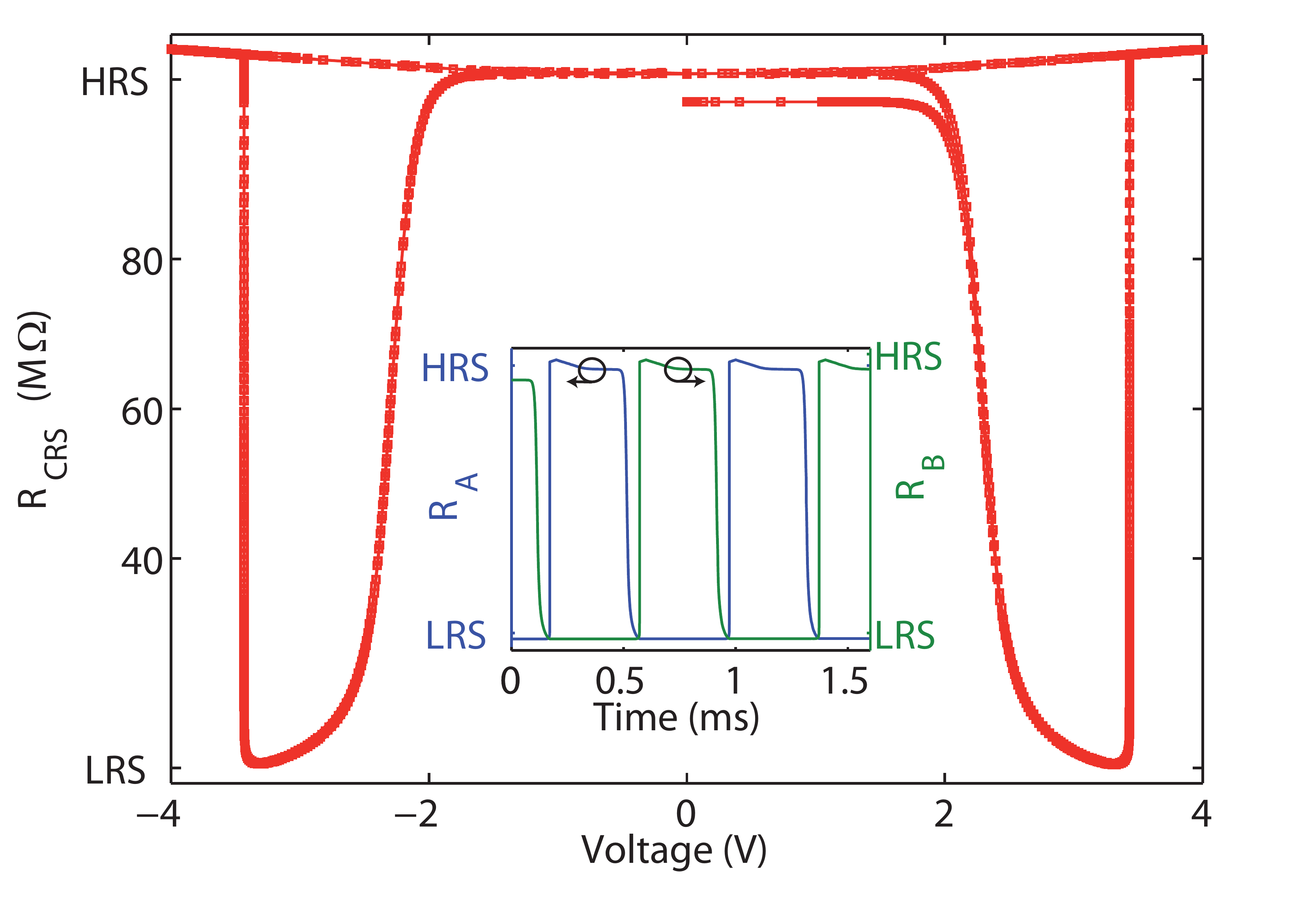}
  \caption{CRS effective resistance for a triangular applied voltage. The inset clearly shows the switching mechanism for memristors A and B. The initial state is slightly less than $R_{\rm OFF}(=R_{\rm HRS}+R_{\rm LRS})$.}\label{fig:crs_r}
\end{figure}

A CRS is a resistive switching device that is built using two memristor devices connected in series with opposite polarities~\cite{crs:linn:2010}. Fig.~\ref{fig:crs_iv}(c) illustrates the modeling results. The figure's inset illustrates a CRS based cross-point array. Each memristor in the figure follows a $I$-$V$ curve relationship that is shown in Fig.~\ref{fig:memristor}(a). The minimum applied voltage for a switch is around $\pm 2.0$~V. Considering the CRS structure as a simple voltage divider, for a LRS/LRS situation\footnote{This situation is defined as the ON state. This is not a stable state so it does not represent a logic state but plays an important role in the switching, READ, and WRITE processes.} minimum $\pm 2$~V is applied across either of the memristors. Please refer to Table~\ref{tab:simparam} for the crossbar memory array parameters. CRS's ON state resistance is $R_{\rm CRS,LRS}=R_{\rm ON}\approx 2R_{\rm LRS}$, where $R_{\rm LRS}$ represents memristor's LRS and CRS's high resistance, $R_{\rm CRS,HRS}=R_{\rm LOGIC}\approx R_{\rm HRS}$, where $R_{\rm HRS}$ indicates memristor's HRS (see Table~\ref{tab:crs_states-trans}). Fig.~\ref{fig:crs_r} highlights the resistance switching of the CRS device. The initial state is programmed to be slightly below $R_{\rm CRS,HRS}$, so there is a difference at the initial curve and the rest of the sweeps. In a memristor device, logic ``0'' and ``1'' are represented with $R_{\rm HRS}$ and $R_{\rm LRS}$, respectively, whereas a CRS device represents logic ``0'' and ``1'' using a combination of low and high resistances which results in overall resistance of $R_{\rm HRS}$ ($R_{\rm OFF}$) for the both logical values.

A fresh CRS device shows a HRS/HRS resistance for memristors A and B. This combination occurs only once (this is not shown in the figure)~\cite{electrochemical:valov:2011}. After applying a positive or negative bias, depending on the polarity of memristors, the device switches to either the ``0" or ``1" state. In Fig.~\ref{fig:crs_iv}(c), red lines are threshold voltages for SET $V_{\rm th,S1}$ and $V_{\rm th,S2}$ and for RESET $V_{\rm th,R1}$ and $V_{\rm th,R2}$. In an ideal CRS device,  $V_{\rm th,SET}=V_{\rm th,S1}=|V_{\rm th,S2}|$ and $V_{\rm th,RESET}=V_{\rm th,R1}=|V_{\rm th,R2}|$. Here $V_{\rm th,SET}=2.4$~V and $V_{\rm th,RESET}=3.6$~V. A successful READ operation occurs if $V_{\rm th,SET}<V_{\rm READ}<V_{\rm th,RESET}$. For a successful WRITE, $V_{\rm th,RESET}<V_{\rm WRITE}$. Consequently, every voltage below $V_{\rm th,SET}$ should not contribute any change in the device state. Possible state transitions are shown in Table~\ref{tab:crs_states-trans}, where $R'$ shows the next resistance state, $R$ illustrates the initial resistance state, and output is a current pulse or spike. In this table, H represents high resistance (either logic states, Logic ``'' or Logic ``1''), and L indicates low resistance.

The simplest analytical model of a CRS can be defined in a relative velocity form, when ${\rm d}w/{\rm d}t={\rm d}w_{\rm A}/{\rm d}t+{\rm d}w_{\rm B}/{\rm d}t$ and the two memristors (A and B) form a voltage divider. Therefore, depending on the combination, the difference between $V_{\rm A}$ and $V_{\rm B}$ can be identified.

\begin{table}[htpb] \centering
    \caption{State transitions in CRS}
    \label{tab:crs_states-trans}

    \begin{small}
    \begin{tabular}{|c|c|c|c|}
    \hline \hline
		{$R$} & {$\Delta V$} & {$R'$} & {Output} \\ \hline
		{H (``1'')} & {$V_{\rm th,S1}<\Delta V<V_{\rm th,R1}$} & {L (ON)} & {pulse} \\ \hline
		{H (``1'')} & {$V_{\rm th,R1}<\Delta V$} & {H (``0'')} & {spike} \\ \hline
		{H (``0'')} & {$V_{\rm th,R2}<\Delta V<V_{\rm th,S2}$} & {L (ON)} & {pulse} \\ \hline
		{H (``0'')} & {$\Delta V<V_{\rm th,R2}$} & {H (``1'')} & {spike} \\ \hline
		{L (ON)} & {$V_{\rm th,R1}<\Delta V$} & {H (``0'')} & {--} \\ \hline
		{L (ON)} & {$\Delta V<V_{\rm th,R2}$} & {H (``1'')} & {--} \\  \hline \hline
		\end{tabular}
    \end{small} 
\end{table}

\begin{figure}[htb!]
\centering
  \includegraphics[width=0.45\textwidth]{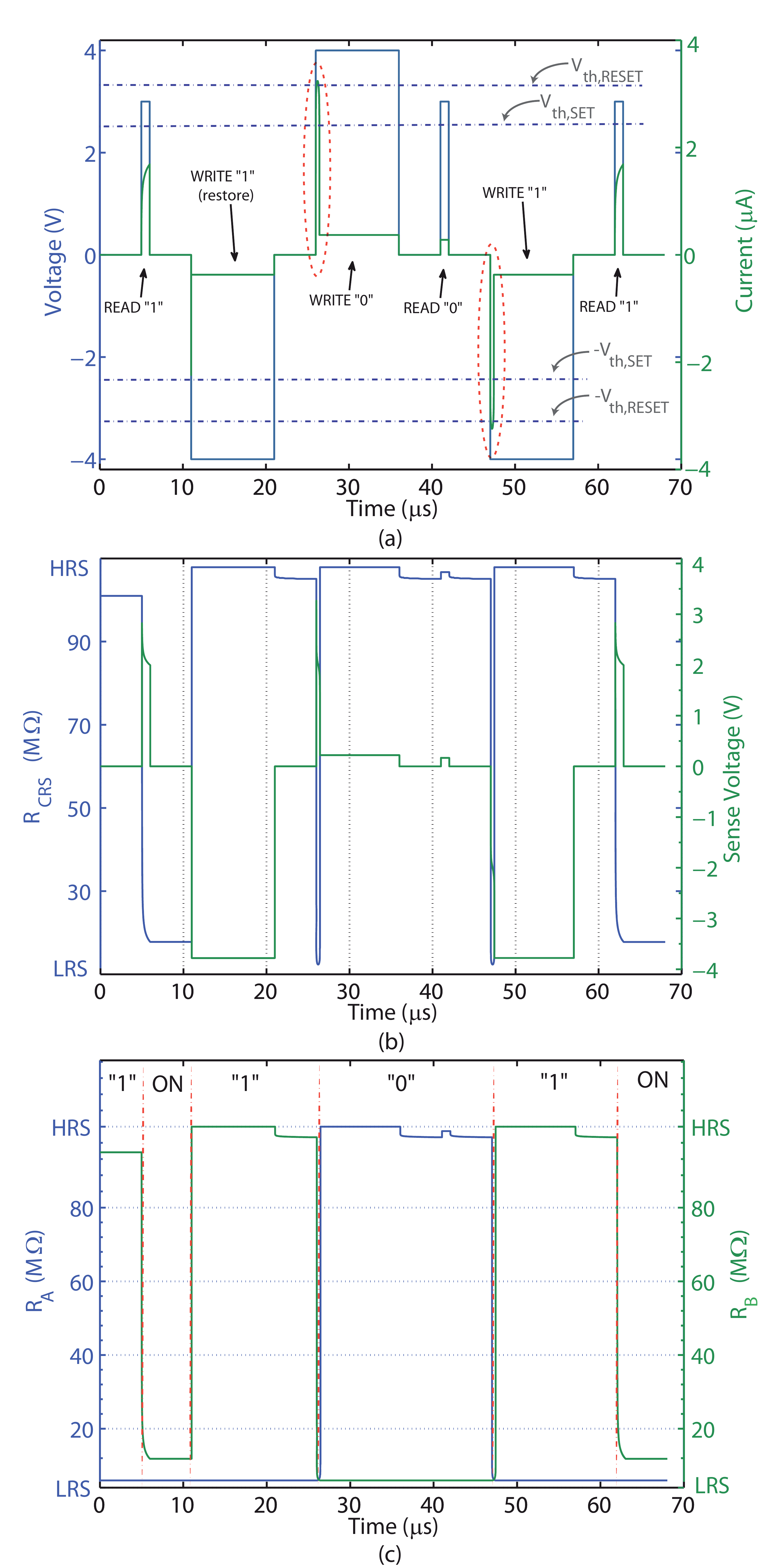}
  \caption{CRS response to a sequence of pulses for READ and WRITE operations. (a) Current response to applied voltage pulses. The dashed lines indicate threshold voltages. The $1~\mu$s READ pulses lie between the $V_{\rm th,SET}$ and the $V_{\rm th,RESET}$. A $5~\mu$s WRITE pulse provides a voltage amplitude beyond $V_{\rm th,RESET}$. (b) Illustrates the total CRS resistance and sense voltage (sensing from middle electrode). As can be seen from the figure, for most of the time $R_{\rm CRS}\approx$HRS. (c) Shows the logic in terms of memristive state for A and B memristors. memristors. Appropriate READ and WRITE pulse widths have been already discussed by Yu~\emph{et al.}~\cite{crs:yu:2010}.}\label{fig:crs_pulse}
\end{figure}

The first feature that appears from the CRS simulation, and device fabrication~\cite{crs:linn:2010, crs:rosezin2011}, is a perfectly symmetric $I$-$V$ curve out of an asymmetric memristor $I$-$V$ curve. The device is programmed initially at logic ``1", LRS/HRS, ($R_{\rm A}\approx$LRS and $R_{\rm B}\approx$HRS). An appropriate READ pulse creates a high potential difference across $R_{\rm A}$ while the voltage difference across $R_{\rm B}$ is not beyond its memristive threshold. Therefore, $R_{\rm A}$ switches to LRS and an ON current (pulse) passes through the CRS device. After a resting time, a negative WRITE pulse is applied to restore ``1", which can be defined as refreshing procedure. Fig.~\ref{fig:crs_pulse}(b) shows that the $R_{\rm CRS}$ settled close to HRS with a logic ``1" stored in the device. Then a positive WRITE pulse tends to write logic ``0", which can be defined as programming step. Depending on the switching speed of the memristors, a short term ON state occurs that causes a relatively large current spike (encircled by red dots in Fig.~\ref{fig:crs_pulse}(a)). There are other functional characteristics that have to be met. For instance, a HRS/HRS state should not appear in any of the situations that are demonstrated by the presented simulation~\cite{crs:yu:2010}. 

The CRS device shows several advantages over a single memristor element for memory applications. This work highlights these advantages for nano crossbar memories.

\section{Crossbar Memory Array}\label{sec:cross}

\begin{figure}[htb!]
\centering
  \includegraphics[width=0.45\textwidth]{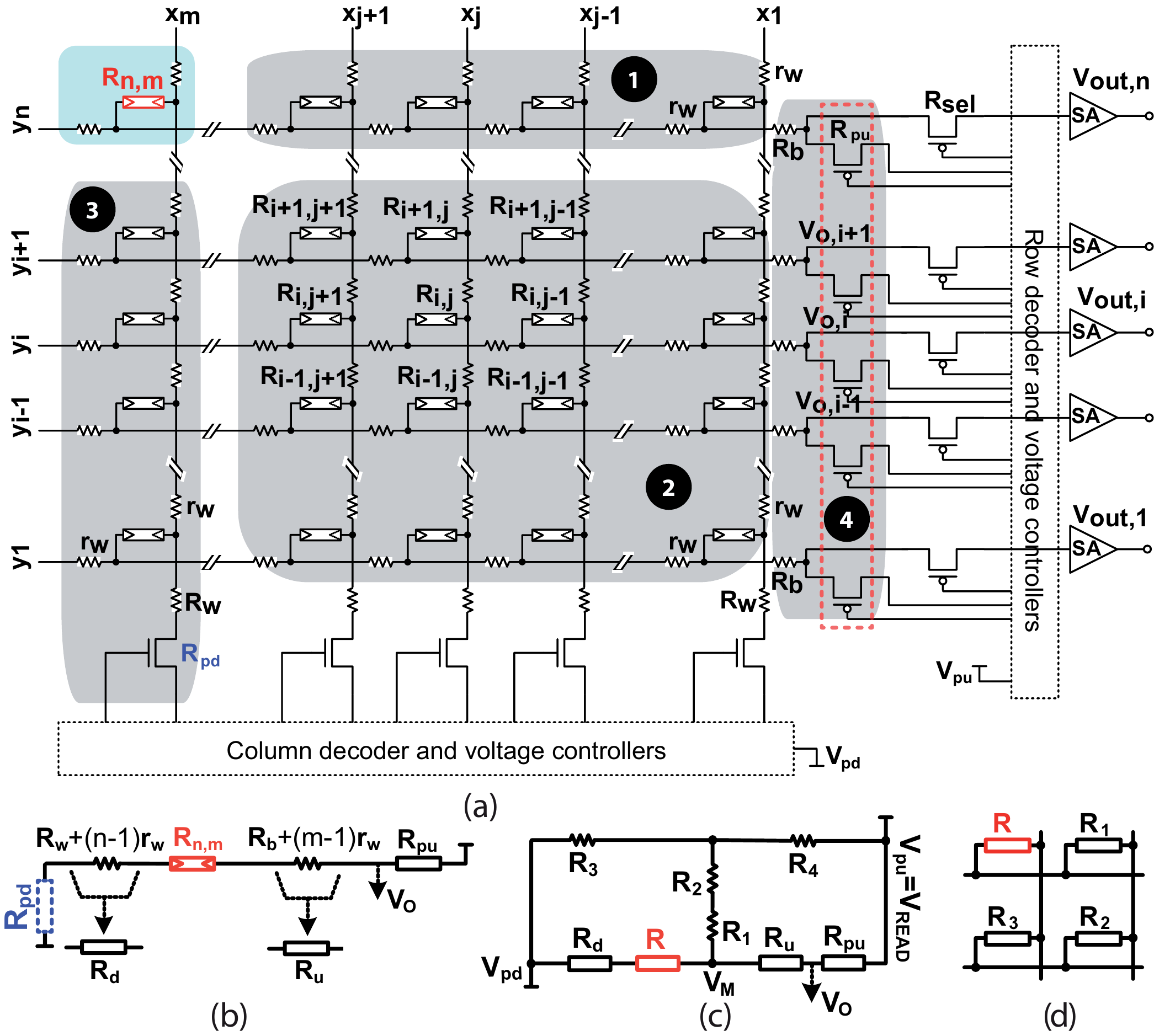}
  \caption{Typical $n\times m$ crossbar array. (a) A hybrid nano/CMOS circuit. Columns show word-lines and rows identify bit-lines. Each $R_{i,j}$ show resistive elements that can be a memristor or a CRS device. Nanowire segment resistance is shown by $r_{\rm w}=R_{\rm line}/n$ (if $n=m$) and the connection between the nanowire and word-line is shown using $R_{\rm w}$. Similarly the bit-line resistance indicated by $R_{\rm b}$. According to the model discussed in Section~\ref{sec:intro}, unit cell resistance of a bit/word line (nanowire) for $F=100$~nm and $n=64$ is around $1~\Omega$ while it is increased to $32~\Omega$ if $F$ reduces to $5~$nm, which is consistent with the unit cell resistance that is reported in~\cite{ITRS:2009} ($1.25~\Omega$). The segment resistance of $r_{\rm w}=1.25~\Omega$ is taken in our simulations. The pull-up resistors (dashed box in red), $R_{\rm pu}$, can be implemented in nano domain and in this paper we assumed them as part of the nano domain. The triangular shape at the output shows a CMOS amplifier that is assumed to have an acceptable sensitivity ($\Delta V$ or $\Delta I$) range of $>100$~mV. While READ process is in progress appropriate signals provided by decoders directs $V_{{\rm o,}i}$ to $V_{{\rm out,}i}$. Stored pattern in groups (1), (2), and (3) can be identifed by $R_{\rm X1}$, $R_{\rm X2}$, and $R_{\rm X3}$, respectively. (b) The ($n^{\rm th}$,$m^{\rm th}$) cell that can be identified in red color is the worst case possible cell for any READ and WRITE schemes. This circuit does not consider sneak-path current. The pull-down resistors, $R_{\rm pd}$ are implemented as part of the CMOS domain and their effects are neglected as the read-out are related to pull-up resistors. (c) Equivalent circuit for the READ scheme with sneak-path and parasitic nanowire resistors considerations. (d) Schematic of a $2\times 2$ array that is a good approximation to the array according to the grouping. }\label{fig:crossbar}
\end{figure}

A crossbar structure as shown in Fig.~\ref{fig:crossbar}(a) is used. This hybrid nano/CMOS implementation is a 3D implementation that the nano domain is stacked on top of the CMOS domain \cite{strukov:2009:four}. The cross-point element could be either a CRS device or a memristor. In order to read any stored bit in $R_{i,j}$, similar to many other reported schemes~\cite{scaling:amsinck:2005, fundamental:flocke:2007, fundamental:flocke:2008, tio2:shin:2011}, here we apply $V_{\rm pu}=V_{\rm READ}$ to the $i^{\rm th}$ bit-line, $j^{\rm th}$ word-line is grounded, and all other word and bit lines are floating. A direct benefit of this approach is the pull-up resistor ($R_{\rm pu}$) can be implemented in nano domain~\cite{fundamental:flocke:2008}. The stored state of the device then can be read by measuring voltage $V_{\rm o}$ that is performed by using CMOS sense amplifiers (SAs). Reading ``1" creates a current pulse, and as a consequence, a voltage pulse appears on the middle electrode. Note that the CRS read-out mechanism does not rely on sensing the middle electrode (ME) and this electrode is floating. The read-out mechanism detects the affect of this current pulse on the bit-line's nanowire capacitor and senses it with the SA array. 

The WRITE scheme that is used is the common accessing method in which the $i^{\rm th}$ bit-line is pulled up, the $j^{\rm th}$ word-line is grounded, and the other lines are all connected to $V_{\rm w}/2$, where $V_{\rm w}$ is the WRITE voltage. This voltage should be high enough to create sufficient voltage difference across the target cell and at the same time having no unwanted affect on the other cells that mainly see a $V_{\rm w}/2$ voltage difference.

Although the WRITE operation in memristors dominates because of its relatively high voltage/current, for CRS device the READ current dominates the current level. Therefore, in our case, the READ operation analysis for energy consumption and performance characteristics is more critical than WRITE operation. Consequently, we discuss the READ and WRITE simulation results here. Table~\ref{tab:simparam} highlights the cross-point junction and array parameters.

\begin{table}[htb!]
  \centering
  \caption{Parameters of the memristive and CRS cross-point junctions and array structure.}
    \begin{tabular}{|c|r|l|l|} \hline \hline
    Parameter 				 & Value 					& Array$^*$ & Description \\ \hline
    $R_{\rm LRS}$  		 & $100$~k$\Omega$ 	& (M) 	& low resistance state \\ \hline		
    $R_{\rm LRS}$  		 & $200$~k$\Omega$ 	& (C) 	& low resistance state \\ \hline		
    $R_{\rm HRS}$ 		 & $100$~M$\Omega$  & (B)		& high resistance state \\ \hline
    $r$ 							 & $10^3$  				& (M)		& resistance ratio \\ \hline		
    $R_{\rm b}$ 			 & $100~\Omega$  	  & (B)		& input resistance of SA$^\dagger$ \\ \hline  
    $R_{\rm w}$ 			 & $100~\Omega$  	  & (B)		& pull-down resistance$^\triangleright$  \\ \hline  
		$R_{\rm pu}$ 			 & $R_{\rm LRS}$ & (B)	  & pull-up resistor$^\ddagger$ \\ \hline			
		$r_{\rm W}$ 			 & $1.25~\Omega/\square$    & (B)	  & parasitic resistor \\ \hline 
		$V_{\rm th,SET}$ 	   & $2.2$~V     		  & (M)	  & SET threshold \\ \hline 
		$|V_{\rm th,RESET}|$ & $1.8$~V     		  & (M)	  & RESET threshold \\ \hline		
		$V_{\rm READ}$ 		   & $1$~V					    & (M)	  & READ voltage \\ \hline		
		$V_{\rm WRITE}$ 	   & $2$~V	  		      & (M)	  & WRITE voltage \\ \hline 
		$|V_{\rm th,SET}|$ 	 & $2.4$~V     		  & (C)	  & SET threshold \\ \hline 
		$|V_{\rm th,RESET}|$ & $3.6$~V     		  & (C)	  & RESET threshold \\ \hline		
		$V_{\rm READ}$ 		   & $2.8$~V	  		    & (C)	  & read voltage \\ \hline	
		$V_{\rm WRITE}$ 	   & $3.8$~V  		      & (C)	  & write voltage \\ \hline	\hline		
    \end{tabular}  \label{tab:simparam}
		\raggedright{\\$^*$ Array type: (M) Memristor-based, (C) CRS-based, (B) Applicable for both.\\
		$^\dagger$ We assumed that bit-lines are directly connected to sense amplifiers (SAs) and there is no $R_{\rm sel}$ in between.\\
		$^\ddagger$ It is assumed that these resistors are implemented in nano domain. Transistors in Fig.~\ref{fig:crossbar} are used to express a more general form of a hybrid nano/CMOS memory.\\
		$^\triangleright$ Lumped parasitic resistance of an activated pull-down.}
\end{table}

To analyze the structure, we need to provide a simplified equivalent circuit for the crossbar structure, as can be seen in Fig.~\ref{fig:crossbar}(b) and (c). Note that $V_{\rm MEM,READ}<V_{\rm CRS,READ}$ and $V_{\rm th,SET}<V_{\rm CRS,READ}<V_{\rm th,RESET}$. Fig.~\ref{fig:crossbar}(c) illustrates the equivalent circuit considering sneak-path currents for the both memristive and CRS-based array. For the sake of simplicity, two series resistors, $R_{\rm 1}$ and $R_{\rm 2}$, are evaluated separately. 
 
The resistor value for the memristive-based and CRS-based circuit can be written as, 
\begin{eqnarray}
R_{1}&=&\frac{R_{\rm X1}}{(m-1)}+r_{\rm w}~,\label{equ:res1}\\
R_{2}&=&\frac{R_{\rm X2}}{(m-1)(n-1)}+r_{\rm w}~,\label{equ:res2}\\
R_{3}&=&\frac{R_{\rm X3}+R_{\rm d}}{(n-1)}~,\label{equ:res3}\\
R_{4}&=&\frac{R_{\rm pu}+R_{\rm u}}{(n-1)}~,\label{equ:res4}
\end{eqnarray}
where $R_{\rm X}$ represents the array's stored pattern in three different groups as seen in Fig.~\ref{fig:crossbar}(a). For the worst case READ or WRITE in a memristive array $R_{\rm X1}=R_{\rm X2}=R_{\rm X3}=R_{\rm LRS}$ and for a CRS array $R_{\rm X1}=R_{\rm X2}\equiv$ logic ``1'' or logic ``0'', whereas $R_{\rm X3}\equiv$ logic ``1''. Although a worst case cell selection is considered here, due to the harmonic series behavior of the overall parallel parasitic resistance, increasing $R_{\rm LRS}$ and/or decreasing the array size, $n=m$, results in better agreement with the analytical approximation for $R_{1}$ and $R_{2}$. These equations then can be used for evaluating the impact of parasitic current paths and parasitic nanowire resistors on the array performance.

The worst case pattern is assumed to be applied when $R_{\rm LRS}\equiv$ logic ``1'' for either memristive or CRS array. In this case, we have the most significant voltage drop because of the both parasitic paths and elements. Therefore, a pattern of either logic ``0" or ``1" for all the elements except those on $j^{\rm th}$ word-line ($x_{j}$) is assumed, which is roughly equivalent to a HRS resistance for a CRS device, $R_{\rm CRS,HRS}=R_{\rm HRS}+R_{\rm LRS}$. A close look at Fig.~\ref{fig:crs_pulse}(a) and (c) shows if the stored logic in all the CRS devices along $x_{j}$, which may or may may not include $R_{i,j}$, is ``1" (LRS/HRS), so applying a $V_{\rm READ}$ can change their states to an ON condition (LRS/LRS). Therefore, during the READ time, $n-1$ devices along $x_{j}$ comprise the ON resistance, $R_{\rm ON}=2R_{\rm LRS}$, in our simulations. Therefore, for a CRS array, $R_{\rm X1}=R_{\rm X2}=R_{\rm HRS}+R_{\rm LRS}$ and $R_{\rm X3}=2R_{\rm LRS}$.

Sizing of the pull-up resistor, $R_{\rm pu}$, as part of the nano domain implementation is a very important factor. For instance, low $R_{\rm LRS}$ devices, e.g. magnetic tunneling junctions (MTJs), interconnection impedance should be also taken into account, whereas in ReRAMs, the LRS resistance is normally $\gtrsim 100~{\rm k}\Omega$ \cite{feedback:yi:2011}. Therefore, our approach provides a generalized analytical form for a nanocrossbar array. In other words, if $R_{\rm LRS}\gg (n+m)r_{\rm W}$, the nanowire overall resistance will be negligible. These considerations along with taking low output potential, $V_{\rm OL}$, and high output potential, $V_{\rm OH}$, lead to an optimal value for $R_{\rm pu}$. We first follow the conventional approach without considering parasitic currents and the nanowire resistors. Note that in a more precise analysis, the sense amplifier's sensitivity is also important to be considered~\cite{memory:zhirnov:2010}. In this situation have,
\begin{eqnarray}
V_{\rm OL}&=&\frac{R_{\rm L}}{R_{\rm L}+R_{\rm pu}}V_{\rm pu}~,\label{equ:voutl}\\
V_{\rm OH}&=&\frac{R_{\rm H}}{R_{\rm H}+R_{\rm pu}}V_{\rm pu}~,\label{equ:vouth}
\end{eqnarray}
where for a memristive array $R_{\rm L}=R_{\rm LRS}$ and $R_{\rm H}=R_{\rm HRS}$, whereas for a CRS array $R_{\rm L}=2R_{\rm LRS}$ and $R_{\rm H}=R_{\rm HRS}+R_{\rm LRS}$. 
These equations are applicable for both of the arrays (memristive and CRS array). Read margin (RM) is defined as $\Delta V=V_{\rm OH}-V_{\rm OL}$. An optimal value of $R_{\rm pu}$ can be extracted from $\partial \Delta V/\partial R_{\rm pu}=0$. For a memristive array, $R_{\rm pu,MEM}=R_{\rm LRS}\sqrt{r}$, where $r=R_{\rm HRS}/R_{\rm LRS}$. Taking parasitic resistors into account (e.g. MJTs), and neglecting sneak-paths, results in $R_{\rm pu,MEM}=\sqrt{rR_{\rm LRS}(R_{\rm LRS}+(n+m)r_{\rm W})}$ as an optimal value for the load resistor. For a CRS array, 
\begin{eqnarray}
R_{\rm pu,CRS}=R_{\rm LRS}\sqrt{2(1+r)}~.\label{equ:Rpucrs}
\end{eqnarray}

A generalized form can be achieved by solving two Kirchhoff's current laws (KCLs) for the equivalent circuit (Fig.~\ref{fig:crossbar}(c)). Therefore, if $R_{12}=R_{1}+R_{2}$ and 
\begin{eqnarray}
x&=&\frac{R_{12}}{R_{4}}+\frac{R_{12}}{R_{3}}-1~,\label{equ:resx}\\
y&=&\frac{1}{R+R_{\rm d}}+\frac{1}{R_{\rm pu}+R_{\rm u}}-\frac{1}{R_{12}}~,\label{equ:resy}\\
V_{\rm M}&=&\frac{R_{12}(xR_{4}-(R_{\rm pu}+R_{\rm u}))}{R_{4}(1+xyR_{12})(R_{\rm pu}+R_{\rm u})}V_{\rm pu}~,\label{equ:vm} \\
V_{\rm O}&=&\frac{R_{\rm u}}{R_{\rm pu}+R_{\rm u}}(V_{\rm pu}-V_{\rm M})~,\label{equ:vo} 
\end{eqnarray}
where $R$ is a memristor or CRS device to be read. This is similar to the ideal condition, $R_{\rm LRS}\Rightarrow V_{\rm OL,MEM}$, $2R_{\rm LRS}\Rightarrow V_{\rm OL,CRS}$, $R_{\rm HRS}\Rightarrow V_{\rm OH,MEM}$, and $R_{\rm HRS}+R_{\rm LRS}\Rightarrow V_{\rm OH,CRS}$. A numerical approach helps designers to identify an optimal value for $R_{\rm pu}$. The significance of this analytical model can be highlighted using a comparison between the optimal values for $R_{\rm pu}$ calculated through Eqs.~(\ref{equ:voutl}) and (\ref{equ:vouth}) and the optimal value calculated via Eq.~(\ref{equ:vo}) (and if parasitic resistors are negligible then via Eq.~(\ref{equ:vm})). The optimal value for a memristive array, neglecting sneak currents and parasitic resistors, is around $3.16~{\rm M}\Omega$ using data from Table~\ref{tab:simparam}. This parameter is a strong function of array size ($n$ and $m$) and $R_{\rm LRS}$. In practical designs, $R_{\rm pu}$ optimal increases as $n~(=m)$ increases. This rate of change can be significantly reduced by a high $R_{\rm LRS}$. This study shows $R_{\rm LRS}>3~{\rm M}\Omega$ results in a significant reduction in dependency of the optimum value to the array size. This analytical approach also indicates that in our case $R_{\rm pu}\approx R_{\rm LRS}$ for CRS and memristive arrays. 

\begin{figure}[htb!]
\centering
  \includegraphics[width=0.3\textwidth]{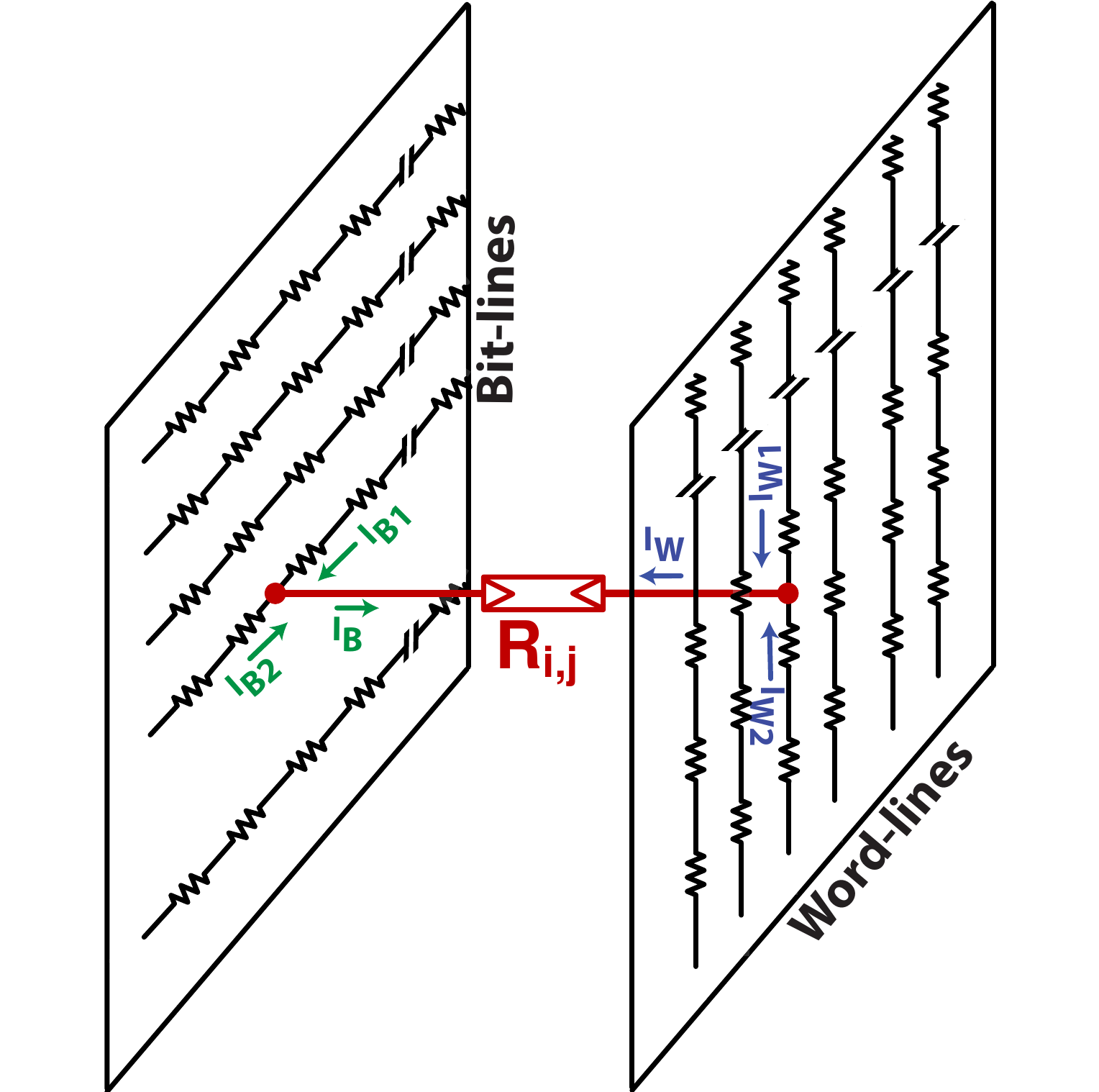}
  \caption{A schematic of an interconnection network in a crossbar array to illustrate the two KCL equations that can be achieved from both bit-line and word-line plates. Generally, current flow through a cross-point device from the $i^{\rm th}$ bit-line is $I_{{\rm B},i}=I_{{\rm B1},i}+I_{{\rm B2},i}$, similarly for the $j^{\rm th}$ word-line is $I_{{\rm W},j}=I_{{\rm W1},j}+I_{{\rm W2},j}$.}\label{fig:kcl}
\end{figure}

In this paper we calculated the voltage pattern using $2mn$ linear equations from a $n\times m$ array. This mathematical framework can be easily implemented using KCL equations in a matrix form. Fig.~\ref{fig:kcl} demonstrates the schematic of how KCL equations work in the two plates. Basically, it shows that for $1<(i~{\rm and}~j)<n$ ($n=m$), 
\begin{eqnarray}
g_{\rm w}V_{{\rm B1},i}+g_{\rm w}V_{{\rm B2},i}&=&G_{i,j}V_{i,j}~,\label{equ:bitline}\\
g_{\rm w}V_{{\rm W1},j}+g_{\rm w}V_{{\rm W2},j}&=&-G_{i,j}V_{i,j}~,\label{equ:wordline}
\end{eqnarray}
where $g_{\rm w}=1/r_{\rm W}$, $G_{i,j}=1/R_{i,j}$, $V_{{\rm B1},i}=V_{{\rm B},i,j+1}-V_{{\rm B},i,j}$, $V_{{\rm B2},i}=V_{{\rm B},i,j-1}-V_{{\rm B},i,j}$, $V_{{\rm W1},j}=V_{{\rm W},i+1,j}-V_{{\rm W},i,j}$, $V_{{\rm W2},j}=V_{{\rm W},i-1,j}-V_{{\rm W},i,j}$, and $V_{i,j}=V_{{\rm B},i,j}-V_{{\rm W},i,j}$. Therefore, we have $2mn$ unknowns and $2mn$ equations. Depending on the READ scheme, the first and last rows (word-line) and columns (bit-line) should be treated differently. Therefore, we can have a control over the voltage pattern and the amount of voltage drop for READ and WRITE operations considering parasitic resistors.

Although, the impact of multiple parasitic currents and nanowire resistors are studied in the literature ~\cite{cross:liang:2010,fundamental:flocke:2007}, a comprehensive analytical approach to address these issues, for the both memristive and CRS arrays, is lacking and it is this issue that is addressed in this paper.

\subsection{Simulations of crossbar array}\label{sub:sec:uncertainty}

Extensive analytical studies have been carried out in the area of nano crossbar memory design~\cite{fundamental:flocke:2007,fundamental:flocke:2008,cross:liang:2010,tio2:shin:2011,scaling:amsinck:2005}. Here we extend these studies and also the comprehensive analytical framework, introduced early in this section, to the simulation of the memristive and CRS-based arrays. 

\begin{figure}[htb!]
\centering
  \includegraphics[width=0.4\textwidth]{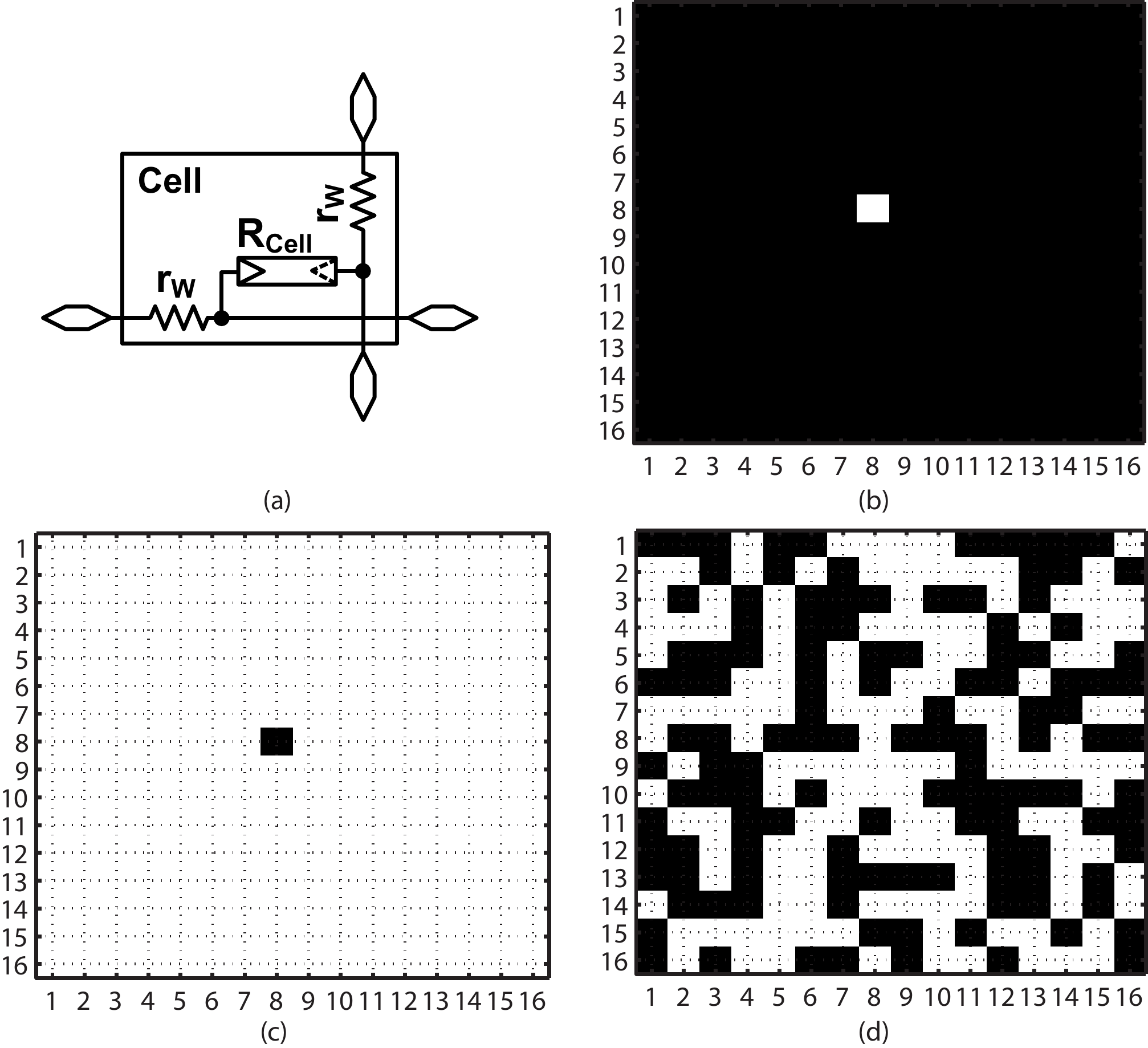}
  \caption{Crossbar memory stored patterns for SPICE simulations. (a) Cross-point unit cell. Horizontal line shows bit-line and vertical wire illustrates word-line. (b) A possible best case in terms sneak-paths for reading $8^{\rm th}$ word-line. We assume a pattern that all the $16$ bits in this word are programmed at their OFF state and there is only one bit with ON state resistance (LRS). The worst case possible is to read from or write in the last word-line. (c) A possible worst case. (d) A random pattern. In all cases, we read the $8^{\rm th}$ column. We initially program one bit with different logic to be able to analyze read margin efficiently. Note that these cases are all relative worst and best cases for comparing the two technologies. A worst case for reading ``1" occurs if selected word line contains only cells with $R_{\rm HRS}$ and the rest of the array are at $R_{\rm LRS}$. Worst case for reading ``0" and writing happen when all the resistive elements are at their $R_{\rm LRS}$ state. For a CRS array, if the array under test is initially programmed to store ``0" for all the cells on selected word line and ``1" for the rest of the array, worst case for reading ``1" occurs, otherwise if the array stores only ``1" logic then the worst case for WRITE operation and reading ``0" occurs.}\label{fig:memsitivecell}
\end{figure}

For simulations in this paper we consider a range of rectangular array sizes from $n=4$,$16$, and $64$ ($n=m$) and three input patterns, the best case, the worst case, and a random pattern for sneak-path current, or interference from neighboring cells, considerations. All patterns are identical for memristor-based and CRS-based arrays. Parameter values can be found in Table~\ref{tab:simparam} and the simulations results are as reported in Table~\ref{tab:result_16x16} for the memory patterns that are given in Fig.~\ref{fig:memsitivecell}. If $\Delta V\ge 100$~mV is acceptable for the CMOS amplifiers, $\Delta V/V_{\rm pu}\ge 10\%$ for memristors and $\Delta V/V_{\rm pu}\ge 3.6\%$ for CRS are acceptable. Only acceptable case for memristor array is the best stored pattern ($16\times 16$) while all the results for a CRS array are acceptable. This study shows that a memristive array needs substantial improvement in the static power dissipation to be an appropriate candidate for the future memory applications, if $R_{\rm LRS}$ values are in the order of k$\Omega$.

\begin{figure}[htb!]
\centering
  \includegraphics[width=0.3\textwidth]{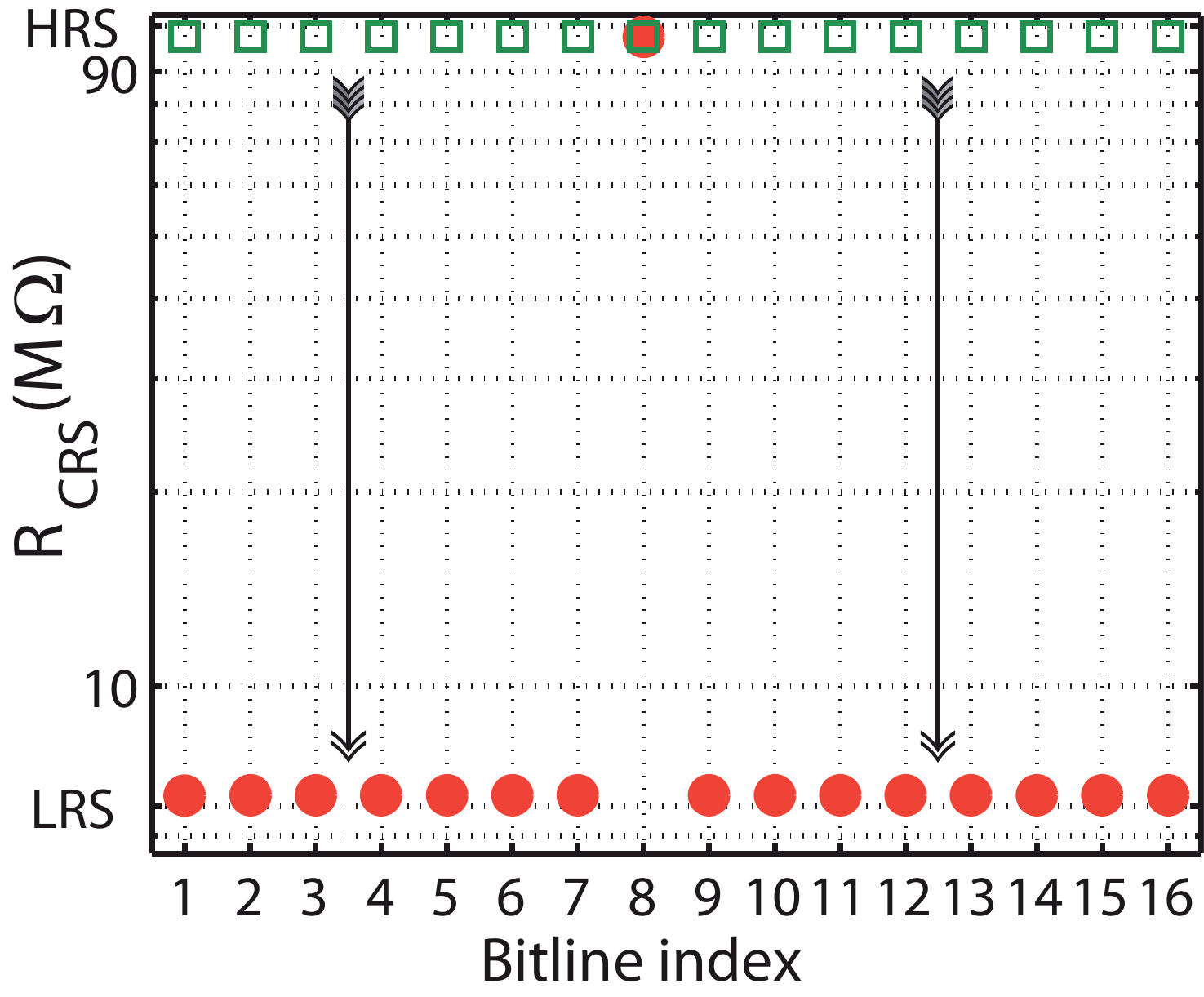}
  \caption{Resistance switch in a column of CRS devices, green rectangular and red circles are the resistance states before and after READ operation, respectively.}\label{fig:crs16x16}
\end{figure}

For a worst case pattern in a CRS-based array, the $(16\times 16)-1$ bits are initially programmed at their LRS/HRS state ($R_{\rm X1}$, $R_{\rm X2}$, and $R_{\rm X3}$), which is effectively equivalent to a HRS state. There is only $1$ bit on the selected word-line that is programmed with a different logic value and this is for reading $V_{\rm OH}$ and $V_{\rm OL}$ at the same time via different bit-lines. Fig.~\ref{fig:crs16x16} illustrates how a worst case happens by applying appropriate READ voltage (here $2.8$V) and $15$ CRS devices switch to their $R_{\rm ON}$ state, which makes a significant difference in terms of the maximum amount of current that can pass through the device. Consequently, this is the main source of power dissipation for a CRS array. A similar scenario is observed for a $64\times 64$ array.  

\begin{figure}[htb!]
\centering
  \includegraphics[width=0.4\textwidth]{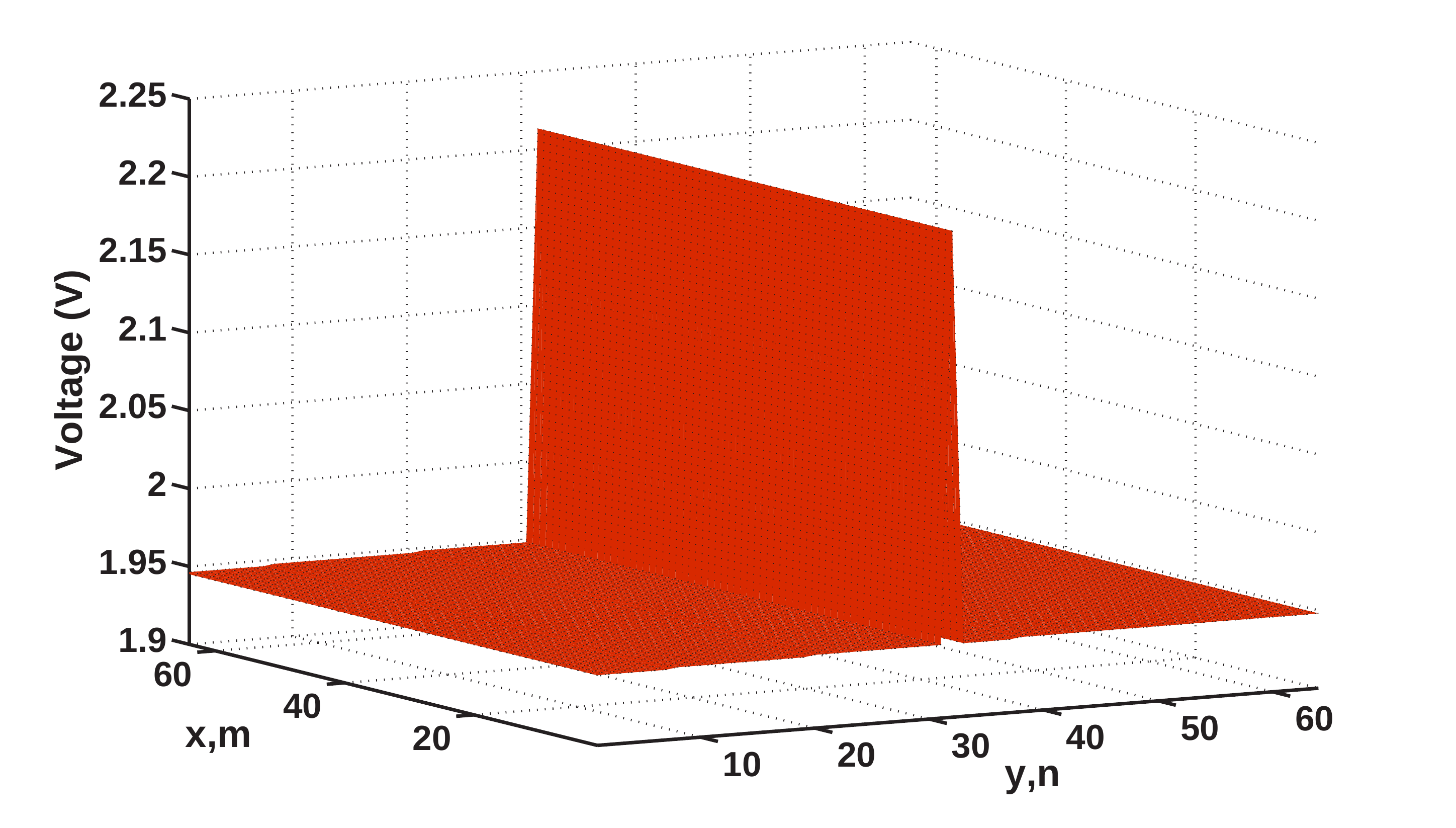}
  \caption{Voltage pattern across a $64\times 64$ CRS cross-point array.}\label{fig:crsvoltpat}
\end{figure}

Fig.~\ref{fig:crsvoltpat} illustrates voltage pattern for a $64\times 64$ CRS array for reading $32^{\rm th}$ word-line. The magnitude of the voltage peak above the settled voltage surface for unselected cells shows a successful READ process. A similar approach can be taken for the WRITE process to show that the minimum requirement (at least the programming threshold voltage) is met on the selected cell(s).  

\begin{table}[htb!]
  \centering
  \caption{Crossbar simulation results}
    \begin{tabular}{|r|r|c|c|r|}  \hline \hline
         Array & \multicolumn{ 2}{|c|}{$\frac{\Delta V}{V_{\rm READ}}$ ($\%$)} & Energy (pJ) & $V_{\rm READ}$ \\ \hline
    \multicolumn{ 5}{|c|}{MEM ($> 10\%$)}            \\ \hline
    \multicolumn{ 1}{|r|}{$16\times 16$} & Worst & \cellcolor{red}$2.3$   & $32.1$ & \multicolumn{ 1}{|c|}{$1$~V} \\ \hline
    \multicolumn{ 1}{|r|}{$16\times 16$} & Best  & \cellcolor{green}$52.6$  & $4.0$ & \multicolumn{ 1}{|c|}{$1$~V} \\ \hline
    \multicolumn{ 1}{|r|}{$16\times 16$} & Random & \cellcolor{red}$5.5$   & $26.9$ & \multicolumn{ 1}{|c|}{$1$~V} \\ \hline
    $64\times 64$ & Worst & \cellcolor{red}$0.45$  & $127.0$ & \multicolumn{ 1}{|c|}{$1$~V} \\ \hline
    \multicolumn{ 5}{|c|}{CRS ($\ge 3.6\%$)}            \\ \hline
    \multicolumn{ 1}{|r|}{$16\times 16$} & Worst & \cellcolor{green}$19.8$  & $119.9$ & \multicolumn{ 1}{|c|}{$2.8$~V} \\ \hline
    \multicolumn{ 1}{|r|}{$16\times 16$} & Best  & \cellcolor{green}$24.3$  & $20.8$ & \multicolumn{ 1}{|c|}{$2.8$~V} \\ \hline
    \multicolumn{ 1}{|r|}{$16\times 16$} & Random & \cellcolor{green}$21.2$  & $89.0$ & \multicolumn{ 1}{|c|}{$2.8$~V} \\ \hline
    $64\times 64$ & Worst & \cellcolor{green}$9.5$  & $483$ & \multicolumn{ 1}{|c|}{$2.8$~V} \\ \hline \hline
    \end{tabular}
  \label{tab:result_16x16}
\end{table}

By analyzing the results for a range of array from $4\times 4$ to $64\times 64$, it is observed that to gain an appropriate and nondestructive READ, $V_{\rm READ}=1$~V, for memristors. This is an intermediate voltage and is low enough to avoid significant change in the device internal state and is high enough to drive a $64\times 64$ array for the last or worst case selected cells. Our simulations indicate that a relatively high LRS ($>3~{\rm M}\Omega$, as reported in~\cite{cross:liang:2010}) guarantee enough read margin as well as sufficient potential across a selective cell for a successful WRITE operation when $r>2$. Therefore, the negative contributions of nanowire parasitic resistors and parasitic (sneak) path currents that are responsible for voltage drop on the selected lines can be both significantly mitigated to a negligible level by increasing $R_{\rm LRS}$ and maintaining $r$ at a level to guarantee a distinguishable high and low state outputs, whether in terms of $\Delta V$ or $\Delta I$.

Endurance requirement\footnote{An endurance of $10^{15}$ is required to replace SRAM and DRAM \cite{ITRS:2009}. There is also an inverse relationship between endurance and data retention. To support reliable large array products, memory technologies must be able to retain data over a long lifetime ($>10$ years at $85\degree$C) with a low defect rates.} in the CRS will be relaxed by utilizing a nondestructive and scalable read-out technique, which significantly reduces the total number of refreshing cycles. A nondestructive readout approach can be found in \cite{tappertzhofen:2011:capacity}.

\subsection{Statistical analysis}

Recent study on memristive switching behavior indicates that there is also a lognormal (long tail) distribution associated with LRS switching (SET)~\cite{feedback:yi:2011}. This certainly reduces the impact of interconnection resistors due to the fact that a significant portion of low resistor states have higher values than the nominal $R_{\rm LRS}$, however, the impact of such distribution on the device switching speed is significant. The lognormal distribution has been also seen in the switching time ReRAMs~\cite{lognormal:medeiros:2011}. However, despite extensive research about the mechanism that causes the lognormal distribution this area is still under intensive discussion~\cite{lognormal:medeiros:2011}.

\begin{figure}[htb!]
\centering
  \includegraphics[width=0.45\textwidth]{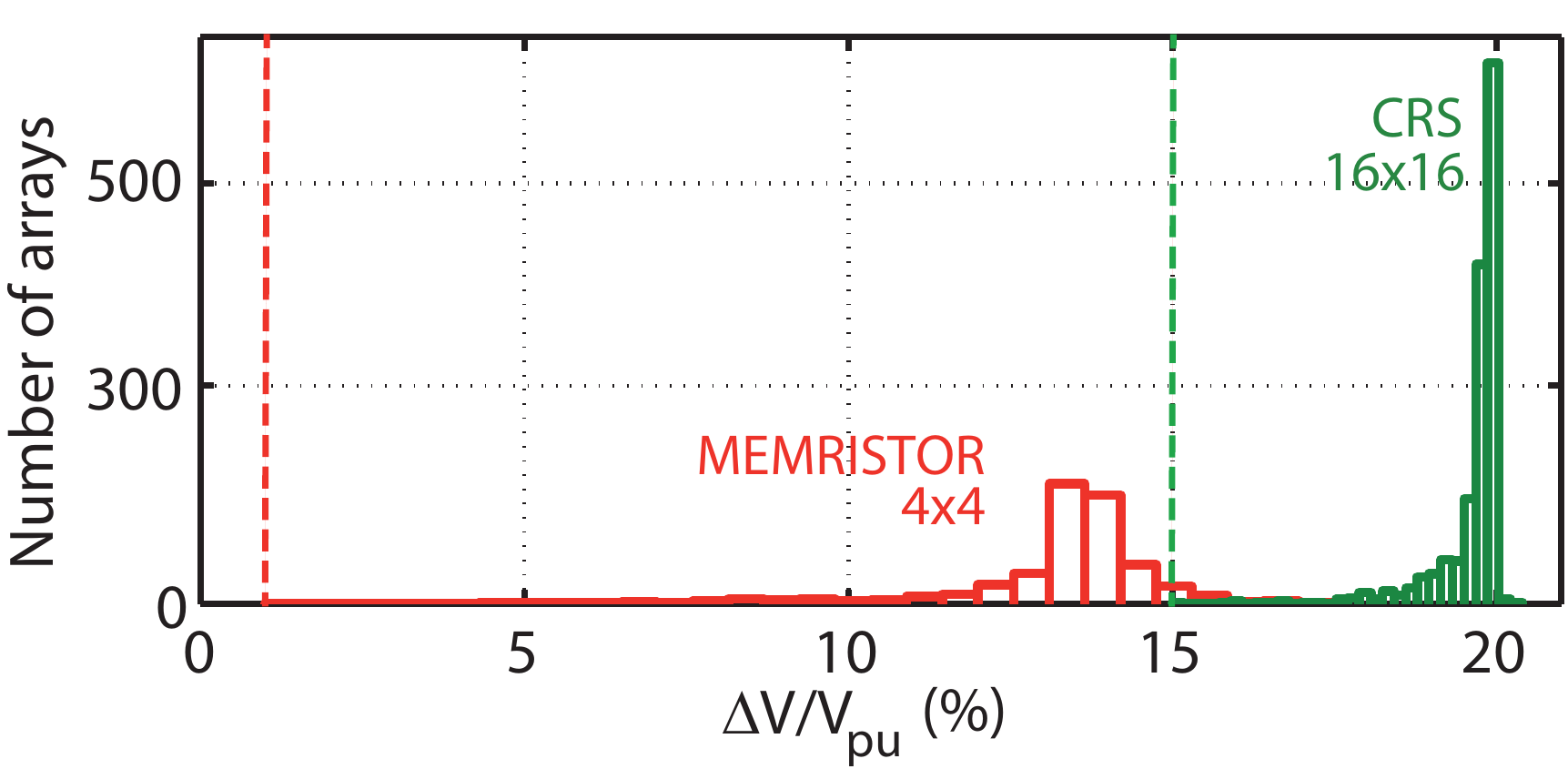}
  \caption{The uncertainty associated with LRS, nanowires process variation, and nonideal initial state programming impact on memristor and CRS array by $1000$ simulation runs, so $1000$ arrays. The red and green lines illustrate minimum read margin for the memristive cross-point and CRS array, respectively.}\label{fig:compare_variation}
\end{figure}

An analysis has been carried out on a $4\times 4$ memristive and a $16\times 16$ CRS arrays through $1000$ Monte Carlo simulations to observe the impact of the uncertainty associated with $R_{\rm LRS}$, device process variation, spatial randomness of the initial state programming, and unfixed applied voltages. The ON state lognormal distribution data is extracted from~\cite{feedback:yi:2011} while a Gaussian distribution is assumed for the the line edge roughness (LER) for devices, nanowires, and variation on the applied voltages. According to~\cite{geometry:hu:2011}, $(-3\sigma,+3\sigma)=(-5.4\%,4.1\%)$ LER and $(-5.5\%,4.8\%)$ thickness fluctuations is assumed for the both $R_{\rm LRS}$ and $R_{\rm HRS}$. We also assumed a normal distribution for initial state programming with $|\pm 3\sigma|=5\%$. Fig.~\ref{fig:compare_variation} demonstrates that the CRS array's output is less sensitive to the overall uncertainty, whereas the memristive array is widely spread out. Minimum value for CRS and memristor arrays are $15\%$ and $1\%$, respectively. As discussed earlier, $\Delta V/V_{\rm pu}\ge 3.6\%$ for the CRS cross-point and $\ge 10\%$ for the memristor array are acceptable. Table~\ref{tab:result_16x16} has already shown that CRS array stored data pattern sensitivity is much ($11$ times) less than the memristive array. Similarly here, while a memristive array read margin is far less than $10\%$, a CRS array guarantee $12\%$ margin. 

In~\cite{geometry:hu:2011} the impact of such variations defined as $R_{\rm XRS}\cdot \theta_{\rm Th}/\theta_{\rm LER}$, where $R_{\rm XRS}$ is either LRS or HRS resistance and the $\theta_{\rm Th}/\theta_{\rm LER}$ define the thickness fluctuations over LER variation. The HRS resistance in a TiO$_2$-based memristor is less affected by the overall variation, whereas LRS variation shows a significant deviation from its nominal value. Likewise, since CRS overall resistance of the (unstable) ON state is $2R_{\rm LRS}$, so it is less affected by such variation.

\subsection{Discussion}\label{sec:sub:dis}

Results of simulations indicate that the most important parameter that should be increased to achieve higher array sizes for a memristive array is $R_{\rm LRS}$. This research verify that a high $r$($=R_{\rm HRS}/R_{\rm LRS}$) does not necessarily improve the substantial amount of parasitic path currents, while a higher $R_{\rm LRS}$ value guarantee a successful READ and WRITE operations. 

In a CRS-based cross-point, the results is more significant since for a high $R_{\rm LRS}$ the effective resistance of HRS and LRS are significantly increased. Assuming a high $R_{\rm LRS}=3~{\rm M}\Omega$, $R_{\rm HRS}=12~{\rm M}\Omega$ ($r=4$), and $V_{\rm READ}=1$~V, READ operation results in a $\Delta V>300$~mV. Analytically, the read margin (RM) does not depend on LRS or HRS resistances and it can be calculated using,
\begin{eqnarray}
{\rm RM}_{\rm MEM}=\frac{1+r-2\sqrt{r}}{r-1}~,\label{equ:rm_mem}
\end{eqnarray}
which means for a $100$~mV limitation $R_{\rm HRS}/R_{\rm LRS}$ ratio should be $\ge 1.3$.
Similarly the simulations results show that such high value assumption for $R_{\rm LRS}$, RM is
\begin{eqnarray}
{\rm RM}_{\rm CRS}=\frac{(r-1)\sqrt{2(1+r)}}{4(1+r)+(3+r)\sqrt{2(1+r)}}~,\label{equ:rm_crs}
\end{eqnarray}
which results $r\ge 1.15$ ($>3.6\%$) minimum requirement for a successful READ. 

Our study also shows that for similar WRITE and READ access time and a high $R_{\rm LRS}$ value, energy dissipation ratio of a CRS array over a memristive array constantly increases. This ratio increases rapidly for $2<r<500$ and (practically) saturates for $500<r<3\times 10^3$. 
The total (static and dynamic) power dissipation for an operation can be calculated using 
\begin{eqnarray}
P_{\rm total}&=&P_{\rm nano}+P_{\rm cmos}~,\label{equ:power:total}\\
P_{\rm nano}&=&P_{\rm sel}+P_{\rm unsel}+P_{\rm pars}~,\label{equ:power:nano}
\end{eqnarray}
where $P_{\rm nano}$, $P_{\rm cmos}$, $P_{\rm sel}$, $P_{\rm unsel}$, and $P_{\rm pars}$ are the nano domain, CMOS domain, selected cells, unselected cells, and parasitic elements (nanowires) power dissipations, respectively. We assume that $P_{\rm cmos}$ for memristive and CRS arrays are comparable and $P_{\rm pars}$ is negligible. The unselected cells power dissipation can be identified by READ and WRITE schemes and can be divided into two (or more) subclasses of half-selected or unselected cells. This approach helps to identify the memory pattern dependency of the total power dissipation assuming a high $R_{\rm LRS}$.

The READ scheme that is discussed earlier, is used as the first READ scheme. Considering such a scheme and high $R_{\rm LRS}$ the total power dissipated by unselected cells (groups (1) and (2) in Fig.~\ref{fig:crossbar}(a), through sneak currents) is negligible. Therefore, the worst case (reading ``1") power consumption for $n\times m$ cells nano in the domain can be calculated through the following equation 
\begin{eqnarray}
P_{\rm nano,MEM}=\frac{nV_{\rm pu,MEM}^2}{R_{\rm LRS}(1+\sqrt{r})}~,\label{equ:power:mem:sch1}
\end{eqnarray}
while similar approach for an $n'\times m'$ CRS array results
\begin{eqnarray}
P_{\rm nano,CRS}=\frac{n'V_{\rm pu,CRS}^2}{R_{\rm LRS}(2+\sqrt{2(1+r)})}~,\label{equ:power:crs:sch1}
\end{eqnarray}
where in this work $V_{\rm pu,MEM}=1$~V and $V_{\rm pu,CRS}=2.8$~V. To fill the gap between the power consumption in memristor and CRS arrays and having similar array size ($n\times m$), $n'=n/c$ and $m'=mc$, where $c$ is a constant that is adjusted to achieve approximately similar power consumption for the two arrays. Here $c=4$, so for example, $1$~K bits of data can be stored either in a $32\times 32$ memristive array or a $8\times 128$ CRS array and have roughly similar power dissipation. 
 
In programming (WRITE) procedure, if $V_{\rm w,MEM}=2$~V and $V_{\rm w,CRS}=3.8$~V, there are $n+m-2$ cells in the both arrays that are half-selected, groups (1) and (3) in Fig.~\ref{fig:crossbar}(a), and $(n-1)(m-1)$ cells that are not selected, ideally $0$~V voltage difference, group (2) in Fig.~\ref{fig:crossbar}(a). The worst case condition is to have all of them at LRS for memristive array and LRS/HRS (logic ``1") for CRS-based crossbar. Therefore, one potential problem with the WRITE scheme is to reset one or more half-selected cross-points. These cells are categorized under unselected cells for the power calculation. Here, there are $1$~V and $1.9$~V potential difference across the half-selected cells in memristor and CRS arrays, respectively, that is sufficiently low to avoid misprogramming. A power consumption analysis for this scheme shows that if $n>16$ and $r>3.5$, the total power dissipated in the CRS nano domain is much lower than the memristive array. The main reason is the all of the half-selected cells have an effective resistance equivalent to $(1+r)R_{\rm LRS}$, whereas the same cells have $R_{\rm LRS}$ that is significantly lower. For example, for $n=100$ and $r=4$ overall improvement in power dissipation is around $38\%$ while for $r=10$ results $70\%$ reduction. This improvement rapidly increases if the number of half-selected cells increases. For instance, a WRITE scheme that activates all bit-lines (pull-up) and $j^{\rm th}$ word-line (grounded) and applying $V_{\rm w}/2$ on the rest of the word lines, can write $n$ bits each time and contains $n(m-1)$ half-selected cells.  

This study indicates that if $R_{\rm LRS}\ll 3~{\rm M}\Omega$, more than $65\%$ and $50\%$ of the total power (and consequently the total energy) is dissipated in half-selected cells during a WRITE operation for the 1-bit WRITE and multi-bit WRITE schemes, respectively. The contribution of half-selected cells is further increased if $R_{\rm LRS}\ge 3~{\rm M}\Omega$. The total power consumption is also rapidly increased as the array size increases. The results also indicate that writing a word (multi-bit) is much more energy efficient than a bit, particularly for CRS-based array.  
Note that for the multi-bit WRITE scheme we applied a two-step WRITE operation (SET-before-RESET) introduced in~\cite{design:xu:2011}. The trade-off between using several WRITE schemes is still an open question.  

Due to the fact that a high $R_{\rm LRS}$ would decrease the energy dissipation and the operation speed at the same time it is very important to note that the nonlinearity of the memristor characteristics plays an important role in identifying the maximum size constraint of a memristive array by identifying the effect of half-selected memory cells. 

Xu~\emph{et al.}~\cite{design:xu:2011} proposed a nonlinearity coefficient to analyze the nonlinearity effect using static resistance values of memristor biased at $V_{\rm w}/p$ and $V_{\rm w}$ as $K_{\rm c}(p,V_{\rm w})=p{R({V_{\rm w}}/{p})}/{R(V_{\rm w})}$. This factor identifies the upper limit for $n$ and $m$ in a memristive array. If parameter $\alpha$ in $I\propto \sinh(\alpha V)$ represents memristor nonlinearity, the factor emphasizes that either higher $\alpha$ or higher $p$ results in a larger $K_{\rm c}(p,V_{\rm w})$. Clearly, the later option is under the designer's control. In fact, this technique effectively creates an intermediate $R_{\rm LRS}$, which is larger than its actual value. Note that the resistance does not necessarily increase if $p$ increases. There is some examples that do not follow similar characteristics, for instance $\sinh^{-1}(\cdot)$ behavior in~\cite{nonvolatile:inoue:2005}. In this case, larger resistances are achievable by decreasing $p$. In~\cite{design:xu:2011}, $K_{\rm c}(p,V_{\rm w})$ is used for bit- and word-lines. Considering the $V_{\rm w}/2$ scheme, this approach is appropriate when $p=2$, $K_{\rm c}(2,V_{\rm w})$. For $p>2$, however, selected word-lines current cannot be calculated with $I(V_{\rm w}/p)$ since the current that passes through an unselected cells on a selected word-line is a function of $V_{\rm w}(1-1/p)$. Hence, calculating an upper limit for $n$ is a function of $K_{\rm r}(p,V_{\rm w})$ that can be defined as,
\begin{eqnarray}
K_{\rm r}(p,V_{\rm w})=\frac{p}{p-1}\frac{R({V_{\rm w}}(1-\frac{1}{p}))}{R(V_{\rm w})}~,\label{equ:power:crs:sch2}
\end{eqnarray}
therefore, a higher nonlinearity coefficient would not necessarily result larger upper limit for memristive array. Furthermore, controlling RESET parameters, such as filament formation process, electrode material, Joule heating process, and TiO$_2$ composition, plays a key role here.  

\section{Conclusion}\label{sec:conclusion}
The presented work provides a foundation and a generic analytical approach to carry out simulations in the design of future memristive-based circuits and systems in general, and CRS arrays in particular. Simulation results indicate that due to sneak-paths and leakage current, a memristive array is faced with a programming and read error rate that aggressively limit the maximum nanocrossbar array size, whereas a CRS array is less affected by these problems. The read margin in a CRS array is also more robust against technology variations such as uncertainty in initial state programming, nanowire process variation, and the associated uncertainty on low resistance state programming.




\end{document}